\documentclass[amsmath,showkeys,showpacs,twocolumn,prl,nofootinbib,longbibliography]{revtex4-1}
\usepackage{amsfonts,amssymb}
\usepackage{graphicx} 
\usepackage[caption=false]{subfig}
\usepackage{bbm}
\usepackage{bm}
\PassOptionsToPackage{unicode}{hyperref}
\PassOptionsToPackage{naturalnames}{hyperref}

\usepackage[hypertexnames=false]{hyperref}

\makeatletter
\newcommand\etal{\emph{et al}\@ifnextchar.{}{.\@}}
\newcommand\etc{etc\@ifnextchar.{}{.\@}}
\newcommand\ie{i.e\@ifnextchar.{}{.\@}}
\newcommand\eg{e.g\@ifnextchar.{}{.\@}}
\makeatother
\def\s{\mathrm{succ}}
\def\f{\mathrm{fail}}
\def\opt{\mathrm{opt}}
\def\loc{\mathrm{loc}}
\def\U{\mathcal{U}}

\def\beq{\begin{equation}}
\def\eeq{\end{equation}}
\def\bsubs{\begin{subequations}}
\def\esubs{\end{subequations}}
\def\bal#1\eal{\begin{align}#1\end{align}}

\begin{document}
\title{Toward an expected utility theory of entanglement:\\ Optimization of a probabilistic entanglement protocol.}
\author{Johnny H.~Esteves}
\author{Antonio Di Lorenzo}
\affiliation{Instituto de F\'{\i}sica, Universidade Federal de Uberl\^{a}ndia, Av. Jo\~{a}o Naves de \'{A}vila 2121, 
Uberl\^{a}ndia, Minas Gerais, 38408-100,  Brazil}
\begin{abstract}
A utility theory of entanglement is formulated, based on the von Neumann--Morgenstern theorem. 
As a specific example, a protocol for measurement--induced entanglement is studied, in which the amount of entanglement obtained depends on the random result of a measurement on 
an ancilla. 
It is shown that the maximization of the utility requires a non--trivial strategy. The results are especially useful when the interaction between the ancilla and the systems to be entangled is weak.
\end{abstract}
\maketitle

The concept of utility was introduced qualitatively by Pascal in his wager argument \cite{Pascal}, and quantitatively by Daniel Bernoulli in order to solve the St. Petersburg paradox \cite{Bernoulli}. 
It was rediscovered by Kelly in connection with the optimization of investments \cite{Kelly}. 
The corresponding criterion was applied by Edward Thorp, Claude and Betty Shannon in their Las Vegas exploit \cite{bankbook}, and is 
still used to determine the optimal amount of wealth to risk in an endeavor, \eg\ when investing in the stock market.  

Entanglement, on the other hand, is a feature unique to quantum mechanics. 
It consists in correlations between two subsystems that cannot be explained by the ignorance of some properties of each subsystem, but on the contrary 
reveal that the very classical notion of each subsystem possessing separated properties is at fault. 
Besides being interesting on a fundamental level \cite{Einstein1935,Schrodinger1935,Bell1964}, entanglement has potential applications. 
For instance, it is essential to produce entangled states on demand, both for quantum computation, quantum teleportation, and quantum cryptography \cite{Bouwmeester2000}. 
Being able to produce entangled particles allows to perform tasks that are not easily performed, or altogether impossible, when using classically correlated systems. 
Entanglement can be quantified for bipartite systems. While there are several possible measures of entanglement (\eg\ entanglement of formation, entanglement cost, and distillable entanglement), all of them must obey some desiderata enunciated by Wootters \cite{Wootters}. 
Entanglement can also be swapped between different systems \cite{swapping}. 
Therefore entanglement is a resource. 

So far, no connection has been made between entanglement and utility, to the best of our knowledge. Here, we shall fill this gap. 
According to the von Neumann--Morgenstern utility theorem \cite{Neumann1947}, a utility function $u(E)$ can be associated to the production of a bipartite system with entanglement $E$. The actual functional form of $u(E)$ will depend on the practical application of entanglement. In general, we can only say that it must be a monotonic (non--decreasing) function of $E$.

Generating entangled particles deterministically, or at least with high efficiency, turns out to be a great challenge. 
It is possible to produce entanglement between two systems by measuring an auxiliary system that has interacted with both (or, equivalently, 
some appropriate degrees of freedom of the systems) \cite{Bose1999,Plenio1999,Hong2002,Duan2003,Sorensen2003}. 
This procedure is known as measurement--induced entanglement or heralded entanglement. It has been observed in atomic systems, in nitrogen vacancies in diamond, and in superconductors \cite{Chou2005,Hofmann2012,Moehring2007,Bernien2013,Roch2014}. 
Due to the intrinsically random nature of measurement in quantum mechanics, measurement--induced entanglement is usually a probabilistic process, \ie\ it succeeds with 
a probability $P_\s<1$ in producing a state $\rho_\s$ with entanglement $E_\s$ (in most entanglement protocols, a success means having created a maximally entangled pair, \ie\ $E_\s=1$, using an appropriate scale for the entanglement measure). In case of failure, the system is left in a state $\rho_\f$ with entanglement $E_\f\le E_\s$. 
Thus, if the failures, which occur with a probability $P_\f=1-P_\s$, are treated as a separate subensemble, the net utility is $N(0) = P_\s u(E_\s)+P_\f u(E_\f)$. 
We shall refer to the strategy of separating the successes and the failures into distinct subensembles as the base strategy. We note that the words `success' and `failure' do not 
necessarily connote positive and negative meaning (although they do for most cases). 

Here, we consider an alternate strategy: We mix a fraction $w$ of the failed attempts to the successful attempts. 
The rate of success of course increases to $P(w)= P_\s+w P_\f$, while the rate of failures decreases to $(1-w) P_\f$. 
The entanglement of the mixture $\rho_1= (\rho_\s + w \rho_\f)/(1+w)$, on the other hand, necessarily decreases compared to the entanglement of $\rho_\s$.  
It is possible to partially compensate for this loss by applying a conditional local operation\footnote{Of course, we cannot consider non--local unitary operations applied to $\rho_\f$, since, if we had at our disposal a non--local operation, we could create entanglement directly without invoking the measurement--induced entanglement protocol.} whenever a failure is  included in the successful attempts. 
The state is thus $\rho_w= (\rho_\s + w U_\loc \rho_\f U^\dagger_\loc)/(1+w)$. In the present work, we individuated the local unitary operations $U_\loc$ that maximize the entanglement of $\rho_w$, then we determined $w_\opt$ the optimal value of $w$ that maximizes the net utility, $C$ being the cost of applying $U_\loc$, 
\beq
N(w)=P(w) u(E_w)+(1-w) P_\f u(E_\f)-w P_\f C .
\eeq
The main findings of our work are: 
(1) For a stronger interaction strength between the ancilla and each subsystem of interest, the increase in the utility from $N(0)$ to $N(w_\opt)$ 
is small, and it is offset by the cost $C$ of applying the unitary local transformation $U_\loc$. 
(2) $N(w_\opt)$ has a finite value in the limit of a vanishing interaction strength between the ancilla and each subsystem of interest.

\begin{figure}[!hb]
\centering
\subfloat[The setup]{\label{fig:setup} 
 \includegraphics[width=0.5\textwidth]{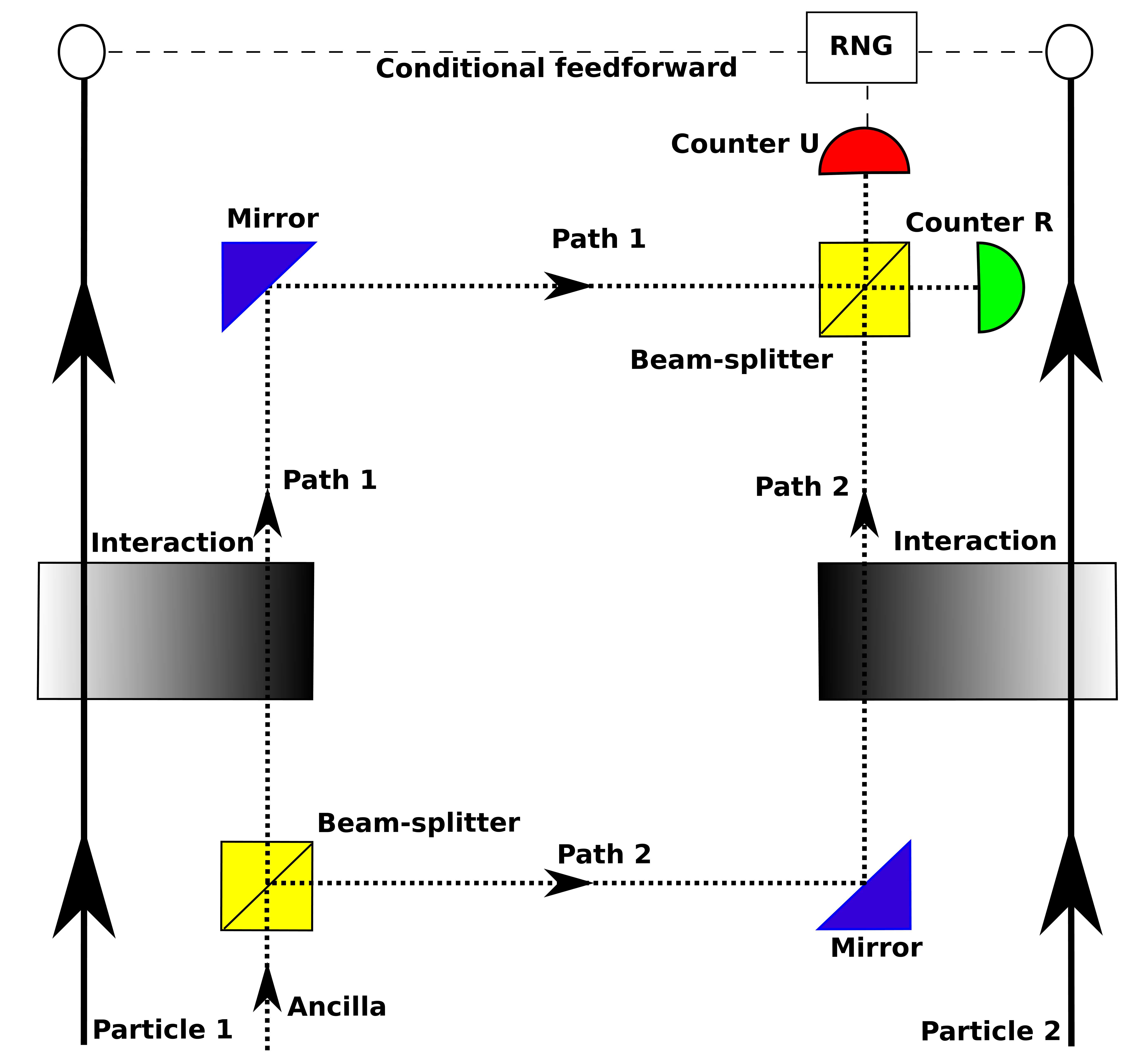}
}\\
\subfloat[The basic block]{\label{fig:inter}
\includegraphics[width=0.4\textwidth]{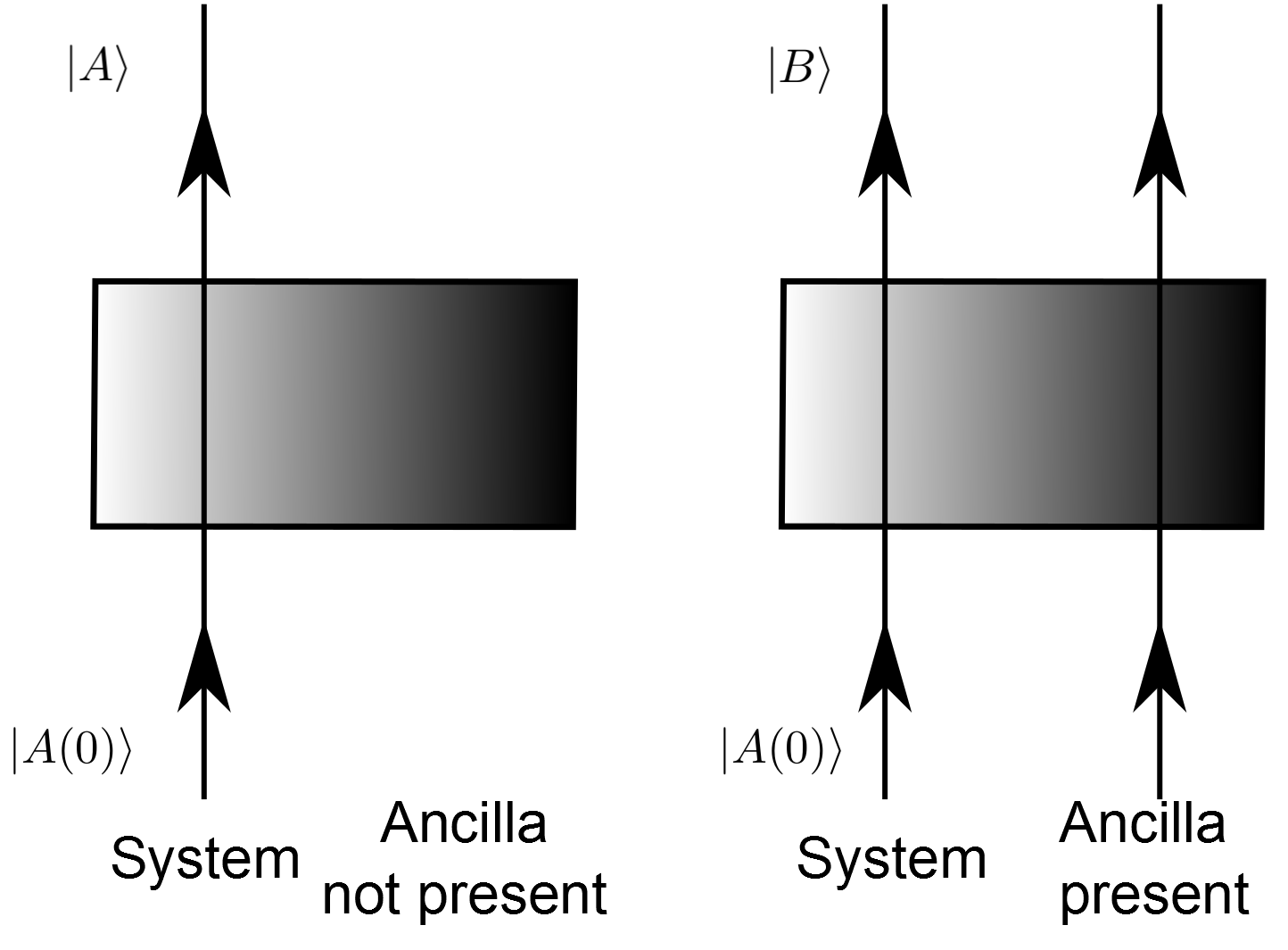}
}
\caption{The setup under consideration. An auxiliary particle, the ancilla, enters a Mach-Zehnder interferometer, where it interacts with the systems to be entangled, labelled 1 and 2, each placed at a different arm, also labelled 1 and 2.}
\end{figure}

%
%
\emph{Setup.} 
The measurement--induced protocol that we consider here is inspired to the one proposed by Haack \etal.\cite{Haack2010} (see also \cite{DiLorenzo2015} for a generalization 
and \cite{DiLorenzo2014b} for a connection with the so--called Quantum Cheshire cat phenomenon) and is described in Fig.~\ref{fig:setup}.  
It requires a Mach--Zehnder interferometer, in which the ancilla is injected, and an interaction between the ancilla and each subsystem of interest. 
We shall assume that the interaction is non--demolition with respect to the spatial degrees of freedom of the ancilla, \ie~if the ancilla enters in a state $|j,\mu_0\rangle$, where 
$j=1,2$ indicates the spatial mode propagating in the $j$--th arm, and $\mu$ is a set of quantum numbers describing its internal degrees of freedom, then the ancilla 
exits the interaction region in the state $|j,\mu\rangle$. 
It is important that the interaction is symmetric in both arms \cite{DiLorenzo2015}. 
After the interaction, the ancilla is detected in either the upper or the right counter. 
If it is detected in $R$, the systems $1$ and $2$ become maximally entangled, for any strength of the interaction (represented by a grayed box). 
If the ancilla is detected in $U$, the systems $1$ and $2$ become only partially entangled, the amount of entanglement depending on the interaction strength. In this case, 
using a random number generator, a fraction $w$ of the systems is retained, to which an appropriate local operation (white circle) is applied so that the entanglement of the mixed state is maximized.
The interaction represented by a grayed rectangle is the fundamental block of the scheme. It is described in Fig.~\ref{fig:inter}. 

In the spirit of von Neumann \cite{vonNeumann1932}, there is no need to describe in detail the interaction through a Hamiltonian, but it suffices to establish the output 
for a set of orthogonal input states. In the present case, the subsystem $j$ is prepared in a state $|A(0)\rangle$. If the ancilla is propagating in the arm $j=1,2$, it interacts with the 
corresponding system, the state of which will change to a state $|B\rangle$ that depends on the interaction, while the state of the other system $\bar{\jmath}=2,1$ changes to 
$|A_{\bar{\jmath}} \rangle$ due to the free evolution. 
For such a dual case scenario, the strength of the interaction can be parametrized with a single, dimensionless real number $\theta\in[0,\pi/2]$, defined by 
\beq |\langle B|A\rangle|^2 = \cos^2(\theta).
\eeq
 Indeed, if the interaction between the subsystem and the ancilla is strong, the subsystem will emerge in a state $|B\rangle$ 
almost orthogonal to $|A\rangle$; if the interaction is weak, instead,  $|A\rangle$ and $|B\rangle$ will be almost indistinguishable.  
We shall use the decomposition $|B\rangle = \cos(\theta)|A\rangle + \sin(\theta)|A^\perp\rangle$, with $|A^\perp\rangle$ orthogonal to $|A\rangle$. 

It is possible to show that, if the ancilla is detected in the right counter, the conditional state of the systems 1 and 2 is the singlet state 
\bal
|\psi\rangle_\s &=\frac{1}{2}\left[ |A_1, B_2\rangle - |B_1, A_2\rangle\right] 
\nonumber \\
&
=\frac{\sin(\theta)}{2}\left[ |A_1, A_2^\perp\rangle -  |A_1^\perp, A_2\rangle \right].
\label{eq:singlet}
\eal
Here, the normalization is such that the square modulus $_\s\langle\psi|\psi\rangle_\s$ equals the probability of success $P_\s = \sin^2(\theta)/2$. 
On the other hand, if the ancilla is detected in the upper counter, which occurs with $P_\f=1-\sin^2(\theta)/2$, the state is 
\bal
|\psi\rangle_\f &=\frac{1}{2}\left[ |A_1, B_2\rangle + |B_1, A_2\rangle\right] 
\nonumber \\
&
=\cos(\theta) |A_1,A_2\rangle+\frac{\sin(\theta)}{2}\left[ |A_1, A_2^\perp\rangle +  |A_1^\perp, A_2\rangle \right].
\label{eq:fail}
\eal
\begin{figure}[h!]
\centering
\includegraphics[width=0.4\textwidth]{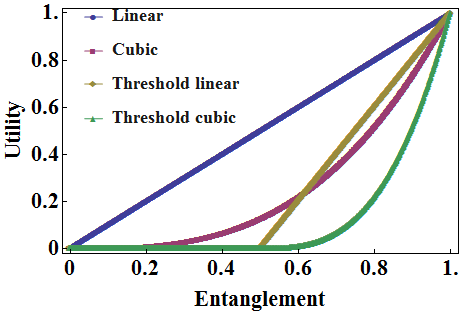}
\caption{\label{fig:grossgain} The utility as a function of entanglement. In this figure we provide four samples of the functions that we used in the rest of the Letter.}
\end{figure}
While the entanglement $E$ of the singlet state $|\psi\rangle_\s$ is always maximal, the entanglement of $|\psi\rangle_\f$ depends on the coupling strength $\theta$. 
For $\theta=\pi/2$ it is $E=1$, \ie\ the protocol always produces a maximally entangled state. In this case, the net utility $N(w)$ is constant in $w$, hence all strategies are equally good. 
However, for $\theta<\pi/2$ the entanglement of $|\psi\rangle_\f$ is less than 1, hence there may be a strategy characterized by a value $w=w_\opt$ that maximizes the net utility $N(w)$. 
\begin{figure}[!t]
\centering
\includegraphics[width=0.45\textwidth]{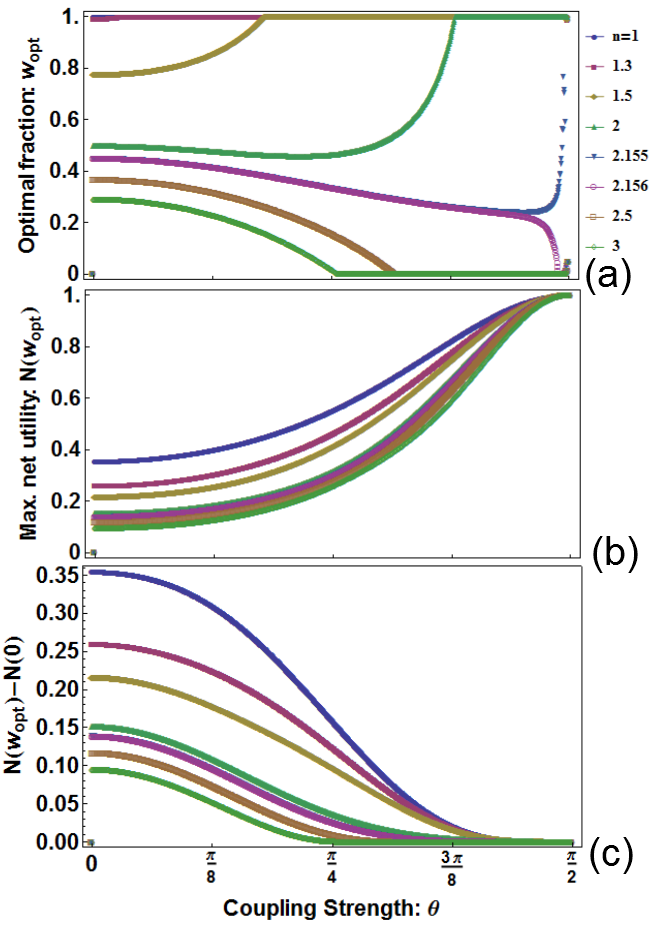}
\caption{\label{fig:optimal1} (a) The optimal fraction of failed attempts to retain, as a function of the coupling strength for the utility functions $u(E)=E^n$ and a cost $C=0$. 
(b) The maximum net utility, corresponding to the $w$ plotted in the subfigure above.
(c) The maximum net utility relative to the utility obtained in the base strategy $N(0)$.}
\end{figure}
\begin{figure}[!t]
\centering
\includegraphics[width=0.45\textwidth]{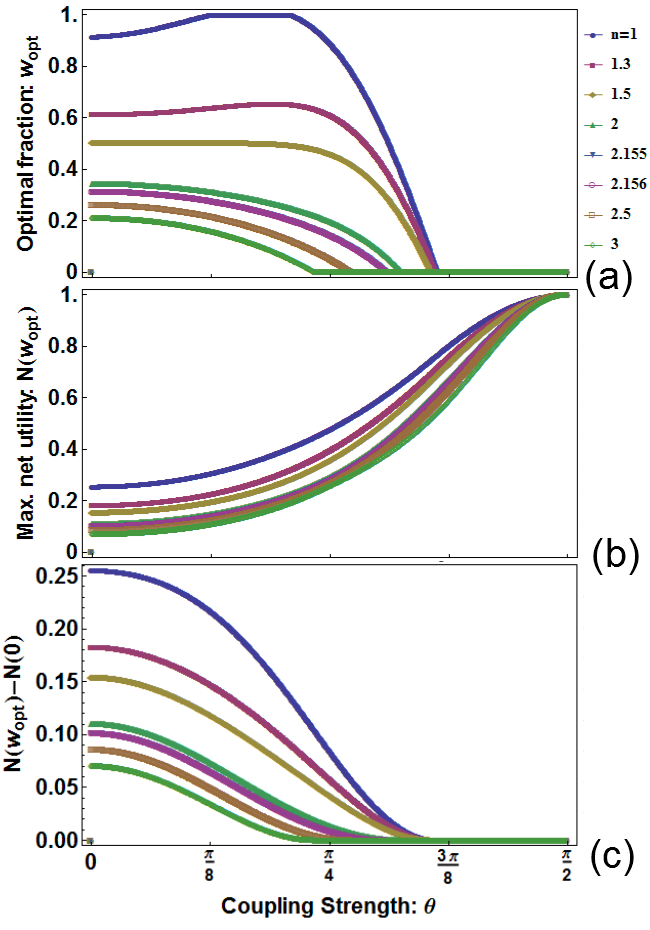}
\caption{\label{fig:optimal2} (a) The optimal fraction of failed attempts to retain, as a function of the coupling strength for the utility functions $u(E)=E^n$ and a cost $C=0.1$. 
(b) The maximum net utility, corresponding to the $w$ plotted in the subfigure above.
(c) The maximum net utility relative to the utility obtained in the base strategy $N(0)$.}
\end{figure}

\emph{Results.} 
In the following, we plot the results for several utility functions, namely powers of entanglement $u(E)=E^n$ and threshold functions 
$u(E)= \Theta(E-E_0) F(E-E_0)$, with $\Theta$ the Heaviside step function and $F$ a monotonically non--decreasing function. 
We considered a utility of the form $u(E)= \Theta(E-E_0) (E-E_0)^n/(1-E_0)^n$, since it may happen that entanglement is not useful below a certain threshold $E_0$. 
We always rescaled the functions $u$ in such a way that the maximum utility is 1. 
\begin{figure}[!h]
\centering
\includegraphics[width=0.45\textwidth]{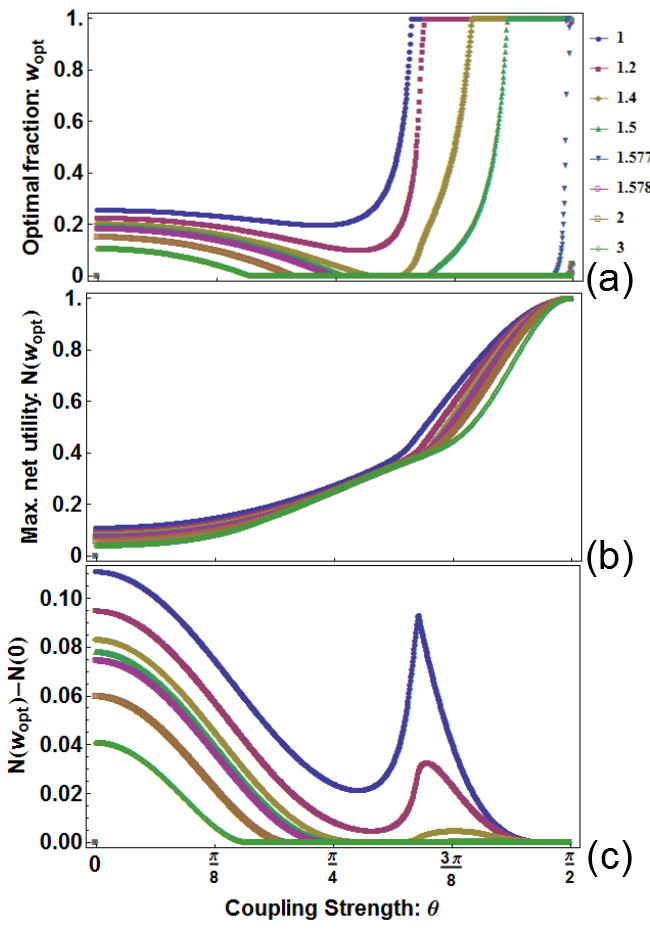}
\caption{\label{fig:optimal3} (a) The optimal fraction of failed attempts to retain, as a function of the coupling strength for the utility functions $u(E)=\Theta(E-0.5) [(E-0.5)/0.5]^n$ and a cost $C=0$. 
(b) The maximum net utility, corresponding to the $w$ plotted in the subfigure above.
(c) The maximum net utility relative to the utility obtained in the base strategy $N(0)$.}
\end{figure}
\begin{figure}[!h]
\centering
\includegraphics[width=0.45\textwidth]{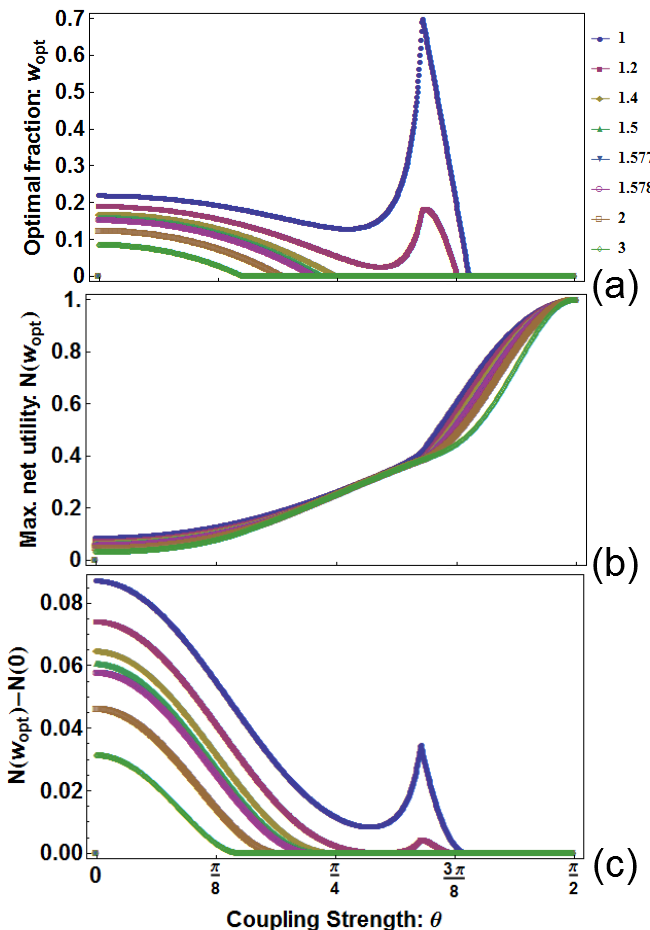}
\caption{\label{fig:optimal4} (a) The optimal fraction of failed attempts to retain, as a function of the coupling strength for the utility functions $u(E)=\Theta(E-0.5) [(E-0.5)/0.5]^n$ and a cost $C=0$. 
(b) The maximum net utility, corresponding to the $w$ plotted in the subfigure above.
(c) The maximum net utility relative to the utility obtained in the base strategy $N(0)$.}
\end{figure}

For a utility of the form $u(E)=E^n$, the optimal strategy turns out to be to always mix $(w_\opt=1$) for any coupling strength, if $n<1.3$. 
 Then, for $1.3<n<2.15$, the optimal strategy consists in keeping all the occurrences ($w_\opt=1$) 
for large enough couplings, while $w_\opt$ decreases for lower couplings. At $n>2.15$, the strategy changes drastically. For large couplings, the best strategy is never to mix, \ie\ the base strategy ($w_\opt=0$), while $w_\opt$ increases for smaller couplings. 
While the strategy changes abruptly with the power $n$, analogously to a phase transition, the net utility has a regular behavior with $n$. 
When including a finite cost for the local unitary operation, the strategy changes sensibly. 
We considered for definiteness a cost $C=0.1$, \ie\ one tenth of the maximum utility. 
Now, at larger coupling strengths and smaller powers $n<2.15$, it is no longer convenient to always mix, but on the contrary one should follow the base strategy. 
This sudden change happens because for larger coupling strengths the state $|\Psi_{1,2}\rangle_U$ is almost fully entangled, so the utility for $\theta\simeq \pi/2$ as a function of 
$w$ is almost degenerate. A finite costs introduces a term $-w P_U C$ that removes such a degeneracy, favoring the strategy of never mixing $w=0$. 
We also considered the family of threshold functions $u(E)=\Theta(E-0.5) [(E-0.5)/0.5]^n$. Indeed in some applications entanglement may have no utility, unless it is above 
a threshold value. Here, we fixed the threshold to $0.5$ for definiteness. When the cost $C$ is zero, we have a behavior similar to that observed for power--law utilities: at some critical value of $n$ the strategy changes abruptly. Furthermore, when we introduce a finite cost, $C=0.1$, this abrupt change is washed out as for the power--law case.

\emph{Method.} 
We proceeded in two steps. First, we found the optimal local unitary transformation $\U_\loc$ such that $E$ is maximized for fixed $w$, and then the optimal fraction $w$ of pairs to which apply the feed--forward, such that $N(w)$ is maximized, given a utility function $u(E)$. 
These two steps were repeated for different values of the coupling strength $\theta$ between the ancilla and each subsystem. 
While in principle we could apply a unitary operation on each subsystem, it can be shown that it suffices to apply a transformation to just one of the two subsystems. 
We recall that a unitary matrix on a two--dimensional Hilbert space can be parametrized as 
$U=
\begin{pmatrix}
 e^{i \beta} \cos (\alpha) & -e^{-i \gamma} \sin (\alpha) \\
 e^{i \gamma} \sin (\alpha) & e^{-i \beta} \cos (\alpha) 
\end{pmatrix}
$. 
Thus, we needed merely to maximize a function of three variables $\alpha,\beta,\gamma$, instead of a function of six variables. 
Operationally, this means that we can apply the feedforward just to one of the particles, rather than to both. 

In more detail, in the first step, we maximized the entanglement of $\rho_w$ using the simplified formula for the concurrence derived in Ref.~\cite{Wang}, then writing an analytic expression for the concurrence with the help of Mathematica, exporting this expression to a Fortran file, and then applying the NLopt libraries \cite{nlopt} in a double loop in 
the variables $\theta$ and $w$, sampling 1000 points for each. The concurrence was then used to find the entanglement of formation \cite{Wootters}. 
The data were saved to a file, then they were analyzed and plotted using again Mathematica for several simple utility functions. 

The second step was relatively straightforward, since we needed to find the maximum of a function $N$ of one variable, $w$. This was done using Mathematica to find the maximum entanglement of formation for fixed $\theta$ and $u(E)$, while $w$ was varying among the 1000 sample points, giving thus a precision of $0.001$ in the location of the optimal value $w_\opt$.

\emph{Discussion.}
While the optimal strategy depends crucially on the form of the utility function, we note some common properties of the net utility as a function of the coupling strength. 
The net utility increases monotonically with the coupling strength.  
Therefore a strong coupling is always preferable, as it leads to maximal entanglement on demand, and as it is more robust agains small asymmetries in the couplings \cite{DiLorenzo2015}. 
On the other hand, for vanishing coupling, the net utility tends to a small but finite value. This signals an essential discontinuity. Indeed, if the coupling were zero from the beginning, there would be no entanglement, and hence no utility. Therefore, the approach proposed in the present Letter may be particularly useful when the coupling strengths available are small. 

\begin{acknowledgments}
\emph{Acknowledgments.} 
This work was performed as part of the Brazilian Instituto Nacional de Ci\^{e}ncia e
Tecnologia para a Informa\c{c}\~{a}o Qu\^{a}ntica (INCT--IQ), it
was supported by CNPq, Conselho Nacional de Desenvolvimento Cient\'{\i}fico e Tecnol\'{o}gico, proc. 311288/2014-6, and by FAPEMIG, Funda\c{c}\~{a}o de Amparo \`{a} Pesquisa de Minas Gerais, proc.~IC-FAPEMIG2016-0269, APQ--01863--14 and PPM-00607-16. 
\end{acknowledgments}

%
%

\begin{thebibliography}{10}
\expandafter\ifx\csname url\endcsname\relax
  \def\url#1{\texttt{#1}}\fi
\expandafter\ifx\csname urlprefix\endcsname\relax\def\urlprefix{; DOI: }\fi
\providecommand{\bibinfo}[2]{#2}
\providecommand{\eprint}[2][]{{#2}}

\bibitem{Pascal}
\bibinfo{author}{Pascal B.} (author), \bibinfo{editor}{Eliot, T.~S.}
 (ed.) \emph{\bibinfo{title}{Pens\'{e}ees}} (\bibinfo{publisher}{E. P. Dutton \& Co., Inc.},
  \bibinfo{address}{New York}, \bibinfo{year}{1958}).

\bibitem{Bernoulli}
 Bernoulli, D. Specimen theori\ae\ nov\ae\ de mensura sortis, \emph{Commentarii Academi\ae Scientiarum Imperialis Petropolitan\ae}, \textbf{V}, 175--192 (1738); 
engl. transl. by Sommer, L. Exposition of a New Theory on the Measurement of Risk, \emph{Econometrica} \textbf{22}, 22--36; DOI:10.2307/1909829 (1954).


\bibitem{Kelly}
 Kelly, J. L.  A New Interpretation of Information Rate. \emph{Bell Syst.~Tech.~J.}~\textbf{35} 917--926; DOI:10.1002/j.1538-7305.1956.tb03809.x (1956).

\bibitem{bankbook}
Poundstone, W. 
\emph{Fortune's Formula: The Untold Story of the Scientific Betting System That Beat the Casinos and Wall Street}, (Hill and Wang, New York, 2006)



\bibitem{Einstein1935}
\bibinfo{author}{Einstein, A.}, \bibinfo{author}{Podolsky, B.} \&
  \bibinfo{author}{Rosen, N.}
\newblock \bibinfo{title}{Can quantum-mechanical description of physical
  reality be considered complete?}
\newblock \emph{\bibinfo{journal}{Phys. Rev.}} \textbf{\bibinfo{volume}{47}},
  \bibinfo{pages}{777--780}\urlprefix{10.1103/PhysRev.47.777} 
 (\bibinfo{year}{1935}).

\bibitem{Schrodinger1935}
\bibinfo{author}{Schr\"{o}dinger, E.}
\newblock \bibinfo{title}{Die gegenw\"{a}rtige {Situation} in der
  {Quantenmechanik}}.
\newblock \emph{\bibinfo{journal}{Naturwissenschaften}}
  \textbf{\bibinfo{volume}{23}}, \bibinfo{pages}{807--812}\urlprefix{10.1007/BF01491891} 
  (\bibinfo{year}{1935}).


\bibitem{Bell1964}
\bibinfo{author}{Bell, J.~S.}
\newblock \bibinfo{title}{On the {Einstein Podolsky Rosen Paradox}}.
\newblock \emph{\bibinfo{journal}{Physics (L.I.)}}
  \textbf{\bibinfo{volume}{1}}, \bibinfo{pages}{195--200}
  (\bibinfo{year}{1964}).
[Reprinted in  \bibinfo{editor}{Bell, M.}, \bibinfo{editor}{Gottfried, K.}  \& \bibinfo{editor}{Veltman, M.} (eds.)
\newblock \emph{\bibinfo{title}{John S. Bell on the Foundations of Quantum Mechanics}}
  (\bibinfo{publisher}{World Scientific}, \bibinfo{address}{Singapore},
  \bibinfo{year}{2001})].

\bibitem{Bouwmeester2000}
\bibinfo{editor}{Bouwmeester, D.}, \bibinfo{editor}{Ekert, A.} \&
  \bibinfo{editor}{Zeilinger, A.} (eds.) \emph{\bibinfo{title}{The Physics of
  Quantum Information: Quantum Cryptography, Quantum Teleportation, Quantum
  Computation}} (\bibinfo{publisher}{Springer Berlin Heidelberg},
  \bibinfo{address}{Berlin}, \bibinfo{year}{2000}).

\bibitem{Wootters} Wootters, W.~K., Entanglement of formation of an arbitrary state
of two qubits. \emph{Phys. Rev. Lett.} 
\textbf{80}, 2245--2248; DOI: 10.1103/PhysRevLett.80.2245
 (1998).

\bibitem{swapping}  Yurke, B. and Stoler, D. Bell's--inequality experiments using independent-particle sources. \emph{Phys. Rev. A} \textbf{46}, 2229--2234 (1992);
\.{Z}ukowski, M., Zeilinger, A., Horne, M. A., and Ekert, A. K. Event-ready-detectors Bell experiment via entanglement swapping. \emph{Phys. Rev. Lett.} \textbf{71}, 4287--4290 (1993). 

\bibitem{Neumann1947}
von Neumann, J. and Morgenstern, O. \emph{Theory of Games and Economic Behavior}, 3rd edition (Princeton University Press, Princeton, 1947). 


\bibitem{Bose1999}
\bibinfo{author}{Bose, S.}, \bibinfo{author}{Knight, P.~L.},
  \bibinfo{author}{Plenio, M.~B.} \& \bibinfo{author}{Vedral, V.}
\newblock \bibinfo{title}{Proposal for teleportation of an atomic state via
  cavity decay}.
\newblock \emph{\bibinfo{journal}{Phys. Rev. Lett.}}
  \textbf{\bibinfo{volume}{83}}, \bibinfo{pages}{5158--5161}\urlprefix{10.1103/PhysRevLett.83.5158}
  (\bibinfo{year}{1999}).


\bibitem{Plenio1999}
\bibinfo{author}{Plenio, M.~B.}, \bibinfo{author}{Huelga, S.~F.},
  \bibinfo{author}{Beige, A.} \& \bibinfo{author}{Knight, P.~L.}
\newblock \bibinfo{title}{Cavity-loss-induced generation of entangled atoms}.
\newblock \emph{\bibinfo{journal}{Phys. Rev. A}} \textbf{\bibinfo{volume}{59}},
  \bibinfo{pages}{2468--2475}\urlprefix{10.1103/PhysRevA.59.2468} (\bibinfo{year}{1999}).


\bibitem{Hong2002}
\bibinfo{author}{Hong, J.} \& \bibinfo{author}{Lee, H.-W.}
\newblock \bibinfo{title}{Quasideterministic generation of entangled atoms in a
  cavity}.
\newblock \emph{\bibinfo{journal}{Phys. Rev. Lett.}}
  \textbf{\bibinfo{volume}{89}}, \bibinfo{pages}{237901}\urlprefix{10.1103/PhysRevLett.89.237901}
  (\bibinfo{year}{2002}).
  

\bibitem{Duan2003}
\bibinfo{author}{Duan, L.-M.} \& \bibinfo{author}{Kimble, H.~J.}
\newblock \bibinfo{title}{Efficient engineering of multiatom entanglement
  through single-photon detections}.
\newblock \emph{\bibinfo{journal}{Phys. Rev. Lett.}}
  \textbf{\bibinfo{volume}{90}}, \bibinfo{pages}{253601}\urlprefix{10.1103/PhysRevLett.90.253601}
  (\bibinfo{year}{2003}).
 

\bibitem{Sorensen2003}
\bibinfo{author}{S\o{}rensen, A.~S.} \& \bibinfo{author}{M\o{}lmer, K.}
\newblock \bibinfo{title}{Measurement induced entanglement and quantum
  computation with atoms in optical cavities}.
\newblock \emph{\bibinfo{journal}{Phys. Rev. Lett.}}
  \textbf{\bibinfo{volume}{91}}, \bibinfo{pages}{097905}\urlprefix{10.1103/PhysRevLett.91.097905}
  (\bibinfo{year}{2003}).


\bibitem{Chou2005}
\bibinfo{author}{Chou, C.~W.} \emph{et~al.}
\newblock \bibinfo{title}{Measurement-induced entanglement for excitation
  stored in remote atomic ensembles}.
\newblock \emph{\bibinfo{journal}{Nature}} \textbf{\bibinfo{volume}{438}},
  \bibinfo{pages}{828--832}\urlprefix{10.1038/nature04353} (\bibinfo{year}{2005}).


\bibitem{Hofmann2012}
\bibinfo{author}{Hofmann, J.} \emph{et~al.}
\newblock \bibinfo{title}{Heralded entanglement between widely separated
  atoms}.
\newblock \emph{\bibinfo{journal}{Science}} \textbf{\bibinfo{volume}{337}},
  \bibinfo{pages}{72--75}\urlprefix{10.1126/science.1221856} (\bibinfo{year}{2012}).


\bibitem{Moehring2007}
\bibinfo{author}{Moehring, D.~L.} \emph{et~al.}
\newblock \bibinfo{title}{Entanglement of single-atom quantum bits at a
  distance}.
\newblock \emph{\bibinfo{journal}{Nature}} \textbf{\bibinfo{volume}{449}},
  \bibinfo{pages}{68--71}\urlprefix{10.1038/nature06118} (\bibinfo{year}{2007}).


\bibitem{Bernien2013}
\bibinfo{author}{Bernien, H.} \emph{et~al.}
\newblock \bibinfo{title}{Heralded entanglement between solid-state qubits
  separated by three metres}.
\newblock \emph{\bibinfo{journal}{Nature}} \textbf{\bibinfo{volume}{497}},
  \bibinfo{pages}{86--90}\urlprefix{10.1038/nature12016} (\bibinfo{year}{2013}).


\bibitem{Roch2014}
\bibinfo{author}{Roch, N.} \emph{et~al.}
\newblock \bibinfo{title}{Observation of measurement-induced entanglement and
  quantum trajectories of remote superconducting qubits}.
\newblock \emph{\bibinfo{journal}{Phys. Rev. Lett.}}
  \textbf{\bibinfo{volume}{112}}, \bibinfo{pages}{170501}\urlprefix{10.1103/PhysRevLett.112.170501}
  (\bibinfo{year}{2014}).

\bibitem{Haack2010}
Haack, G. , F\"{o}rster, H., and B\"{u}ttiker,. M. Parity detection and entanglement
with a Mach-Zehnder interferometer. \emph{Phys. Rev. A} \textbf{82} 155303 (2010).

\bibitem{DiLorenzo2015}
Di Lorenzo, A. Post-selection induced deterministic and probabilistic entanglement with strong and weak interactions, \emph{submitted to Ann. Phys. (NY)}. 

\bibitem{DiLorenzo2014b}
\bibinfo{author}{Di~Lorenzo, A.}
\newblock \bibinfo{title}{Postselection-induced entanglement swapping from a
  vacuum-excitation entangled state to separate quantum systems}.
\newblock \emph{\bibinfo{journal}{Phys. Rev. A}} \textbf{\bibinfo{volume}{90}},
  \bibinfo{pages}{022121}\urlprefix{10.1103/PhysRevA.90.022121} (\bibinfo{year}{2014}).


\bibitem{vonNeumann1932}
\bibinfo{author}{von Neumann, J.}
\newblock \emph{\bibinfo{title}{Mathematische Grundlagen der Quantenmechanik}}
  (\bibinfo{publisher}{Springer}, \bibinfo{address}{Berlin},
  \bibinfo{year}{1932}).
\newblock \bibinfo{note}{[\emph{Mathematical Foundations of Quantum Mechanics}
  (Princeton University Press, Princeton, 1996)]}.

\bibitem{Wang} Wang, A.--M., A Simplified and Obvious Expression of Concurrence in
Wootters' Measure of Entanglement of a Pair of Qubits. \emph{Chin. Phys. Lett.} \textbf{20} 1907--1909 (2003). 

\bibitem{nlopt}
Steven G. Johnson, The NLopt nonlinear-optimization package, http://ab-initio.mit.edu/nlopt; 
Krister Svanberg, "A class of globally convergent optimization methods based on conservative convex separable approximations", SIAM J. Optim. 12 (2), p. 555--573 (2002).

\end{thebibliography}

\end{document}